\documentstyle[aps]{revtex}\newcommand{\be}{\begin{equation}}

\newcommand{\ee}{\end{equation}}
\newcommand{\bea}{\begin{eqnarray}}
\newcommand{\eea}{\end{eqnarray}}
\newcommand{\nn}{\nonumber \\}

\font\mybb=msbm10 at 10pt
\def\bb#1{\hbox{\mybb#1}}
\def\bZ {\bb{Z}}
\def\bR {\bb{R}}
\def\bE {\bb{E}}

\def\tr{\rm tr}

\begin{document}
 \draft
\twocolumn[\hsize\textwidth\columnwidth\hsize\csname
@twocolumnfalse\endcsname

\title{Black Holes and Calogero Models}
\author{G.W. Gibbons and {}~P.K. Townsend }
\address{
DAMTP, Univ. of Cambridge, Silver St., Cambridge CB3 9EW,  UK\\
 }
\maketitle
\begin{abstract}
We argue that the large $n$ limit of the $n$-particle $SU(1,1|2)$
superconformal Calogero model provides a microscopic description of
the extreme Reissner-Nordstr{\"o}m black hole in the
near-horizon limit.
\end{abstract}
\vskip2pc]

An amusing feature of the matrix model approach to M-theory is that
apparently intractable problems of quantum D=11 supergravity are resolved by a 
return to {\sl non-relativistic} quantum mechanics \cite{BFSS}. The M(atrix)
model Hamiltonian, which can be viewed as that of an $SU(\infty)$ D=1
super-Yang-Mills (SYM) theory, was originally found from a light-front
gauge-fixed version of  the D=11 supermembrane \cite{dhn}, but the
interpretation given to it in \cite{BFSS} was inspired by the observation
\cite{pkt} that there is a close similarity between this supermembrane 
Hamiltonian and
that of $n$ IIA D0-branes \cite{wit} in the large $n$ limit. However, because
the D0-brane Hamiltonian differs from that of the M(atrix) model by the
inclusion of relativistic corrections, the precise connection has only recently
become clear. By viewing the limit in which relativistic corrections are
supressed as one in which the spacelike circle of $S^1$-compactified M-theory
becomes null, it was shown in \cite{ss} that all degrees of freedom
other than D0-branes  are also supressed, thus establishing that the M(atrix)
model Hamiltonian captures the full dynamics of (uncompactified)
M-theory. Here we wish to observe, because it will provide a 
useful perspective on our later
results, that this limit can be understood as a `near-horizon' limit. 
In the `dual-frame' metric \cite{dgt}, the D0-brane 
solution of IIA supergravity approaches $adS_2\times S^8$ \cite{bst}. 
Near the $adS_2$ Killing
horizon, and for sufficiently large $n$, there is a dual description of the
D0-brane dynamics \cite{maldatwo} in terms of IIA supergravity on $adS_2\times
S^8$ \cite{bst} (which is the null reduction of the M-wave \cite{BBPT}).  

Because $adS_2$ has an $SL(2;\bR)$ isometry group one might expect the
M(atrix) model Hamiltonian to exhibit this symmetry as a worldline 
conformal symmetry, in
analogy to the dual adS/CFT descriptions of D3-brane dynamics 
\cite{malda} in
a similar `near-horizon' limit \cite{GT}. The SYM interpretation of 
the M(atrix)
model makes it clear that there can be no such symmetry because D-dimensional 
SYM theories are conformally invariant only for $D=4$. In accord with this
observation, the $SL(2;\bR)$ symmetry of the $adS_2\times S^8$ metric does not
extend to the complete supergravity solution because it is broken by
the dilaton
field. Nevertheless, since the value of the dilaton is simply related to the
matrix model coupling constant one might expect there to be a `generalized'
conformal symmetry taking a matrix model at one value of the coupling constant
to the same model at a different value of the coupling constant. Just such a
`generalized' conformal symmetry was exhibited in \cite{yoneya}, but the above
explanation of this result indicates that it should be a general phenomenon
applicable not only to D0-branes but to all 0-brane intersections for which the
dual-frame supergravity metric has an $adS_2$ factor. The `generalized'
$SL(2;\bR)$ conformal invariance will become a genuine conformal invariance in
those cases for which the dilaton is constant. These `genuinely 
conformal' cases are the focus of this paper.

A case in point is the four intersecting D3-brane configuration  
$$
\begin{array}{lccccccccc}
D3: & 1 & 2 & 3 & - & - & - & - & - & - \nn
D3: & 1 & - & - & 4 & 5 & - & - & - & - \nn
D3: & - & 2 & - & 4 & - & 6 & - & - & - \nn
D3: & - & - & 3 & - & 5 & 6 & - & - & - 
\end{array}
$$
As explained in \cite{tsey}, this corresponds to a black hole solution of the
$T^6$ reduction of IIB supergravity for which all scalar fields, including the
dilaton, approach constant values near the horizon. The special case for which
the scalar fields are everywhere constant is the extreme Reissner-Nordstr{\"o}m
(RN) black hole. In all cases the near horizon geometry is $adS_2\times S^2$  
and the isometry supergroup is $SU(1,1|2)$. Assuming that $adS_2\times S^2$ is
equally a solution of the full, non-perturbative, IIB superstring
theory (which seems likely in view of the results of \cite{kallosh})
we may conclude that $SU(1,1|2)$ acts as a superconformal group on the
quantum mechanical model governing the fluctuation of the branes in the region
of the intersection. But what is this model? Whatever it is, we
expect that it has a dual description as $N=2$ D=4 supergravity on 
$adS_2\times S^2$ in a limit that involves interpreting each of the 
four supergravity D3-branes as a large number of coincident
microscopic D3-branes. Equivalently, we expect some limit of the 
superconformal mechanics model to provide a microscopic description of
the extreme RN black hole, at least near the horizon. The
determination of this model is therefore likely to be an important
step in our understanding of the quantum mechanics of black holes. 

The field theory on the string intersection of any two of the four
D3-branes of the above configuration is a (4,4) supersymmetric D=2 SYM
theory. The intersection with the remaining two D3-branes must reduce 
this to a D=1 N=4 superconformal quantum mechanics, so we 
need some $n$-particle $SU(1,1|2)$-invariant superconformal 
mechanics that is related, in the large $n$ limit, to a reduction of 
a D=2 SYM theory. One such model, omitting fermions, is the $n$ particle   
Calogero model \cite{cal} with Hamiltonian
\be\label{calogero}
H = {1\over 2} \sum_i p_i^2 +  \sum_{i< j} {\lambda^2 \over
(q_i-q_j)^2}
\ee 
where $(p_i,q_i)$ ($i=1,\dots,n$) are the n-particle phase space 
coordinates, and $\lambda$ is a coupling constant\footnote{Calogero 
models have arisen previously in the context of the matrix model describing
D0-brane dynamics \cite{poly} but since the latter cannot be conformal
invariant, for the reasons given above, they are necessarily different from
those considered here.}. This model, like a number of variants of it, is
integrable and has been studied extensively in this context (see e.g.
\cite{perelomov}). A distinguishing feature of the particular model
defined by the above  Hamiltonian is that its action is invariant 
under an $SL(2;\bR)$ group acting as a worldline conformal 
group \cite{regge}, so we shall call it the `Conformal Calogero' (CC) 
model. The conformal symmetry arises from an action of $SL(2;\bR)$
on phase space, generated by $(H,K,D)$ where $H$ is the Hamiltonian
(\ref{calogero}), $K= {1\over2}\sum_i q^2_i$ generates conformal
boosts and $D= -{1\over2}\sum_i p_iq_i$ generates dilatations. The generators
$H$ and $K$ are lightlike with respect to the Killing form on
$sl(2;\bR)$ while $D$ is spacelike. In our conventions timelike
generators are `compact'. 

We propose that the model describing the 0-brane intersection of the above four
D3-brane configuration is the N=4 supersymmetric extension of the CC 
 model, and hence that this model provides a microscopic
description of the extreme RN black hole in the near-horizon limit.
The N=2 superconformal Calogero model is described by the superpotential 
$\lambda \sum_{i<j}\log |q_i-q_j|$. It is a special case of the N=2
supersymmetric, but generically non-conformal, Calogero-Moser (CM) model
studied in \cite{dizzy}; this fact will
play a part in our proposal. Unfortunately, the N=4 supersymmetric extensions 
of  the CC and CM models have not yet been constructed but the main features 
are clear and we shall discuss them later. First we wish to explain the
motivation for our proposal in the context of the bosonic model.  

The $n$-particle CC model was shown in \cite{GN} to  be equivalent in the large
$n$ limit to a D=2 $SU(n)$  gauge theory on a cylinder. A related observation
that we shall elaborate on here is that the CC model can be obtained
by symplectic reduction of a class of matrix models. Consider, for
example, 
the space of hermitian $n\times n$ matrices $X$
with the flat metric $\tr (dX)(d\bar X)$. The corresponding free 
particle mechanics
model is manifestly conformal invariant. It is  also invariant under $SU(n)$
transformations acting by conjugation on $X$. The corresponding conserved
`angular momenta' are encoded in the  conserved traceless hermitian matrix
\be
\mu = i[X,P]\, ,
\ee
where $P={\dot X}$ is the momentum canonically conjugate to $X$. The 
idea now is
to work at some fixed values of the angular momenta. This gives 
constraints, and
one quotients phase space by the action generated by these constraints
to get a reduced Hamiltonian system on the quotient. In order
to obtain the CC system this way one must choose the angular
momenta such that $\mu$ has $n-1$ equal eigenvalues $\lambda$. The 
stability group of the matrix 
$\mu$ is then $S(U(1)\times U(n-1)) \subset U(n)$, and the action of 
this group may be
used to bring $X$ to the diagonal form $X={\rm diag} (q_1,\dots,q_n)$ 
and $P$ to
a form with diagonal entries $p_i$ and off-diagonal entries 
$P_{ij}= i\lambda /(q_i-q_j)$. The reduced Hamiltonian, ${1\over2}\tr P^2$,
is just (\ref{calogero}).

It will prove instructive to consider the $n=2$ case in more detail.
In this case the configuration space is $\bE^4$. We can write $X$ as
\be\label{param}
X=U\Delta U^{-1},
\ee
where $\Delta=diag(q_1,q_2)$ and $U$ is the $SU(2)$ matrix
\be
U = \pmatrix{u & v \cr -\bar v & \bar u} \qquad (|u|^2+|v|^2=1).
\ee
There is a $U(1)$  redundancy in this description because we can take
$(u,\bar v) \rightarrow e^{i\alpha}(u, \bar v)$ without changing $X$.
We thus have a parametrization of $X$ in terms of $(q_1,q_2)$ and the 
coordinates of $SU(2)/U(1)\cong S^2$. Introducing the centre of mass coordinate
$Q=(q_1+q_2)/2$ and the relative position coordinate $q=(q_1-q_2)$ we find that
\be\label{relative}
ds^2 = 2 dQ^2 + {1\over2}\left[dq^2 + q^2 d\Omega_2^2\right].
\ee
Note that the 3-metric describing the relative motion on $\bE^3$ is
flat. The angular momentum matrix
$\mu$ has eigenvalues $\pm\lambda$ where $\lambda$ is the length of the angular
momentum 3-vector ${\bf L}$ associated with the $SO(3)$ isometry of $S^2$. At
fixed $\lambda$, the Hamiltonian is 
\be\label{ham}
H= {1\over2}\left[ p^2 + {g\over q^2}\right]\, ,
\ee
where 
\be\label{discrep}
g= \lambda^2\, .
\ee
This is the conformal mechanics Hamiltonian of de Alfaro, Fubini and Furlan 
(DFF) \cite{DFF} with coupling constant $g$. The CC models thus provide a 
natural generalization of  DFF conformal mechanics.

The $N=2$ supersymmetric extension of the DFF model, with $SU(1,1|1)$ 
superconformal symmetry was constructed in \cite{akulov}. This model
is clearly related to the 2-particle N=2 superconformal Calogero 
model in the same
way as above. The superpotential in the latter case is
\be 
\lambda \log |q_1-q_2| = \lambda \log |q|\, ,
\ee
but this is precisely the superpotential of N=2 superconformal
mechanics. Conversely, by simply adding in a trivial centre of mass
motion one can obtain the 2-particle N=2 superconformal Calogero model
by a simple change of variables, and the $n$-particle model is a
straightforward generalization \cite{dizzy}. The $N=4$ extension of 
the DFF model, with $SU(1,1|2)$ superconformal symmetry, was
constructed in \cite{ivanov}; the construction is not completely obvious
because there is still only one physical boson variable, $q$, despite
the N=4 supersymmetry. However, it should be clear from the above
discussion that the N=4 2-particle superconformal Calogero model is
already implicit in the N=4 superconformal mechanics. We expect that
the $n$-particle generalization will again be straightforward, but the
complete construction will not be attempted here. 

We now come to the main motivation for our proposal. It was shown in
\cite{kaletal} that the N=4 superconformal mechanics model describes the
dynamics of a superparticle of unit mass (and charge to mass 
ratio equal to that
of the black hole) in the near-horizon geometry of an extreme RN black hole in
the limit of large black hole mass\footnote{Solutions $\Phi$ of the
covariant Klein-Gordan equation on $adS_2$ in satisfy
$i\partial_t\Phi = \hat H \Phi$ where $2\hat H= qp(q^{-1}p) + 
g^2/q^2$, and the inner product is $\int dr r^{-1}|\Phi|^2$.
It follows that $\hat H$ provides the correct resolution of the
operator ordering ambiguity inherent in the classical hamiltonian
(\ref{ham}). This ordering differs from the 'naive' one used in
studies of quantum conformal mechanics but the change of ordering 
does not affect the qualitative properties of the model needed here.}. 
The coupling constant was found to be 
\be\label{relcoupling}
g= 4\ell^2\, ,
\ee
where $\ell$ is the particle's orbital angular momentum operator quantum
number. Actually, the result of \cite{kaletal} is that $g=4L^2$ where $L^2$ is
the operator with eigenvalues $\ell(\ell+1)$, but the shift of $\ell^2$ to
$\ell(\ell+1)$ can be interpreted as a consequence of integrating out the
fermions \cite{akulov,az}, so one finds $g=4\ell^2$ if the fermions
are simply omitted. Even so, there is an apparent discrepancy with
(\ref{discrep}) because it would be natural to suppose from our derivation of
conformal mechanics from the $2\times 2$ hermitian matrix model that, in the
quantum theory, $\lambda$ should take on integer values, but this is consistent
with (\ref{relcoupling}), and integer $\ell$, only if $\lambda$ 
is an {\sl even} integer. The resolution of this discrepancy is 
that in arriving
at (\ref{ham}) we implicity assumed that $q$ was positive. If one allows $q$ to
be negative, then the parametrization (\ref{param}) of $X$ has an additional
redundancy which can be removed by identification of antipodal points on the
2-sphere. Only those harmonics with even angular momentum quantum
number are well defined on $S^2/\bZ_2 \cong \bR P^2$, so allowing $q$ to be
negative leads to the restriction $\lambda=2\ell$ for integer $\ell$.

The derivation of conformal mechanics from a charged particle on $adS_2\times
S^2$ makes crucial use of a coordinate system for which the $adS_2$ metric is
\be
ds^2= - R^4 q^{-4} dt^2 + 4R^2 q^{-2} dq^2\, ,
\ee
where $R$ is the radius of curvature.
The singularity at $q=\infty$ is just a coordinate singularity at a 
Killing horizon of $h=\partial_t$. The vector fields 
\be
d= t\partial_t + {1\over2}q\partial_q\, , \qquad
k= \left(t^2 + q^4/R^2\right)\partial_t  + tq\partial_q\, .
\ee
are also Killing, and $(h,d,k)$ have the $sl(2;\bR)$ commutation relations
\be\label{alg}
[d,h]=-h\, ,\quad [d,k]=k \, ,\qquad [h,k] =2d\, .
\ee
The key result of \cite{kaletal} is that the particle trajectory $q(t)$ is
described by a relativistic conformal mechanics, and that the Hamiltonian
describing this evolution in $t$ reduces in the large $R$ limit to the
non-relativistic Hamiltonian (\ref{ham}). Just as this Hamiltonian,
$H$, is associated with the Killing vector field $h$, so there are two
other functions on phase space, $D$ and $K$ associated to $d$ and $k$.
Their Poisson bracket algebra is isomorphic to (\ref{alg}). 

In the quantum theory, the DFF Hamiltonian has a continuous spectrum with $E>0$
but no ground state at $E=0$ \cite{DFF}. This feature is a reflection of the
incompleteness of the classical dynamics because a state of zero energy would
be time-independent and hence associated with a fixed set of $\partial_t$ on
$adS_2$, i.e. its Killing horizon. The classical cure for the
incompleteness due to the Killing horizon is simply to choose a global
coordinate system on $adS_2$, e.g.
\be
t = {\sin \tau \over \cos \tau  -\sin\rho}\, ,
\qquad q^2 = {R\cos\rho \over \cos \tau  -\sin\rho}\, ,
\ee
where $|\rho|<\pi/2$, and $\tau$ is periodically identified with period $2\pi$.
The metric is now 
\be
ds^2_2 = (R\sec\rho)^2[-d\tau^2 + d\rho^2] \, .
\ee
We see that $\partial_\tau$ is Killing. It is a compact generator of
$SL(2;\bR)$; in fact
\be\label{comb}
\partial_\tau = {1\over2}(h+k)\, .
\ee
Classical evolution in $\tau$ is complete because $\partial_\tau$ has no
horizon. The corresponding Hamiltonian $\tilde H(\rho,p_\rho)$ can be
found by solving the mass-shell constraint in the new coordinates, but
the quantum states it evolves belong to a new Hilbert space. It is
also no longer obvious how to take the
non-relativistic limit and hence unclear how to generalize to the
multi-particle case. For these reasons we shall not pursue this
approach here. Instead, we retain the original Hilbert space but
evolve the states via a new, non-conformal, Hamiltonian $H'= H+K$, 
as suggested by the formula (\ref{comb}). This was also done in \cite{DFF},
where it was found that $H'$ has a unique ground state with a 
series of evenly spaced excited states, as expected for
evolution in a periodically identified time parameter. Curiously, despite the
fact that orbits of $\partial_\tau$ pass through $q=\infty$, a particle in the
ground state of $H'$ has zero probability of being at $q=\infty$
because the ground state wave-function is \cite{DFF}
\be\label{gs}
\psi_0 = q^\alpha e^{-q^2/4}\, , \qquad (\alpha = {1\over2}\left[ 1 +
\sqrt{1+4g}\right])\, .
\ee

Two particles do not make a black hole, so we now wish to generalize the above
$n=2$ discussion to $n>2$.  A complication of the $n>2$ case is that there are
many possibilities for the angular-momentum matrix $\mu$. Only the choice
described above, $n-1$ equal eigenvalues, leads to the n-particle conformal
Calogero model. A partial motivation for this choice comes from the 
observation \cite{az} that the one independent eigenvalue $\lambda$ of the $n=2$
case is an  `almost-central' charge in the superconformal algebra; it is not
truly central because it can be removed by a redefinition of the $sl(2;\bR)$
charges but, for a given definition, the anticommutator of the odd charges is
$\lambda$-dependent. The equal eigenvalue condition for $n>2$ is thus analogous
to a BPS condition. In any case, having made this choice we arrive at the CC
Hamiltonian (\ref{calogero}), which we now rewrite as
\be
H = H_{n-1} + {1\over2} p_n^2 + {\lambda^2\over q_n^2}\sum_{i=1}^{n-1} 
\left(1-q_i/q_n\right)^{-2}\, .
\ee
where $H_{n-1}$ contains all terms independent of $(q_n,p_n)$.
If we now suppose that all $q_i$ are small except $q_n$ (so that we may omit
${\cal O}(q_i/q_n)$ terms), and take $n$ large (so that we can ignore
reduced-mass effects from factoring out the centre of mass motion) then the
Hamiltonian governing the motion of the $n$th particle is the DFF model with
\be\label{couple}
g= 2(n-1)\lambda^2 = 8(n-1)\ell^2\, , \qquad (n\gg1).
\ee
Recalling that in the quantum theory $\ell$ is an integer multiple 
of Planck's constant, we see that the large $n$ limit is one in which
the particle orbiting the cluster of $n-1$ particles acquires a
macroscopic angular momentum.

We thus arrive at a picture of an extreme black hole as a composite of a large 
number of particles interacting via a repulsive inverse cube force
law. However, this picture is rather misleading because, as mentioned above, 
the variable $q$ is actually an {\sl inverse} radial variable in the
sense that $q=\infty$ corresponds to the black hole horizon. The cluster of
$n-1$ particles near $q=0$ is actually much further from the horizon than the
one at large $q$, although they are  still in the `near-horizon' region of the
black hole. One can view them as living at the $adS_2$ boundary, and the large
$n$ CC model as the boundary conformal field theory in the sense of the adS/CFT
correspondence \cite{malda}. It is then natural to interpret these `Calogerons'
as the microscopic degrees  of freedom of the black hole.

The CC model describes the dynamics of $n$ {\sl ordered} particles in the sense
that if we  choose the $q_i$ such that $q_{i+1}-q_i>0$ then the CC dynamics
implemented by $H$ preserves this ordering. As for conformal
mechanics, the full dynamics in which we allow $q_{i+1}-q_i\le0$ will
be described by some other Hamiltonian. If we follow the DFF approach
described above we would replace the CC Hamiltonian by the new
Hamiltonian $H_{CM}= H+\omega^2K$ for some constant $\omega$. Now
\be
K \equiv {1\over 2}\sum_i q_i^2 = {1\over 2n} \sum_{i<j} (q_i-q_j)^2
+ {n\over 2}Q^2
\ee
where $Q= (1/n)\sum_i q_i$ is the centre of mass position. We may set
$Q=0$ without loss of generality, in which case
\be\label{calmoser}
H_{CM}= {1\over2}\sum_i p_i^2 +  \sum_{i<j}{\lambda^2\over 
(q_i-q_j)^2} + {\omega^2\over 2n} \sum_{i<j}(q_i-q_j)^2 
\ee
This is the Calogero-Moser (CM) Hamiltonian; it has 
a unique ground state and towers of excited states with energy
spacings $2\omega,3\omega,\dots, n\omega$ \cite{cal,dizzy}. It is
therefore natural to associate $H_{CM}$ with a time parameter that is 
periodically identified with period \footnote{In the $n=2$ case the
periodicity could be taken to be $\pi\omega^{-1}$ but the 
inclusion of fermions will require a periodicity of $2\pi\omega^{-1}$
even in this case. One way to see this is to note that
$adS_2$ can be embedded in $\bE^{(2,1)}$, in which case its Killing
spinors are the restriction of constant spinors on $\bE^{(2,1)}$.
It follows that the Killing spinors of $adS_2$ are antiperiodic in $\tau$
when written in a spin basis associated to the one-forms $d\tau, d\rho$,
where $\tau$ and $\rho$ are the $adS_2$ coordinates introduced
previously.} $2\pi\omega^{-1}$.

The ground state wave-function of $H_{CM}$
is similar to (\ref{gs}) in that
particles in the ground state have zero probability of being at any of the
boundaries between the $2^{n-1}$ regions with different signs of the relative
coordinates $q_{i+1}-q_i$. Because classical trajectories connect these
regions it is natural to consider them as distinct, but in the quantum theory
we must choose to put the particles in one region. It is thus natural to assign
the black hole an entropy $S=(n-1)\log 2$ (cf. \cite{BM}). On the 
other hand, the n-particle Calogero-Moser Hamiltonian describes a system of 
size $L\sim \sqrt{n}$ as $n\rightarrow \infty$ for fixed $\omega$ and
$\lambda$ \cite{dizzy}, so the identification of this
system with the RN black hole implies that the area $A$ of the black
hole horizon is $A\sim L^2$. It follows that $S\sim A$, in qualitative
agreement with the Bekenstein-Hawking entropy.


\end{document}